\documentclass
[showpacs,letterpaper,prl,amsfonts,amssymb,oneside,10pt,preprint,balancelastpage,notitlepage,twocolumn]{revtex4}%
\usepackage{amsfonts}
\usepackage{amsmath}
\usepackage{amssymb}
\usepackage{graphicx}%
\begin{document}
\title{Quantum squeezing of optical dissipative structures}
\author{Isabel P\'{e}rez--Arjona$^{1,2}$, Eugenio Rold\'{a}n$^{1}$, and Germ\'{a}n J.
de Valc\'{a}rcel$^{1}$,}
\affiliation{$^{1}$Departament d'\`{O}ptica, Universitat de Val\`{e}ncia, Dr. Moliner 50,
46100--Burjassot, Spain}
\affiliation{$^{2}$Departament de F\'{\i}sica Aplicada, Escola Polit\`{e}cnica Superior de
Gandia, Universitat Polit\`{e}cnica de Val\`{e}ncia -- Ctra. Nazaret--Oliva
S/N, ,46730--Grau de Gandia, Spain}

\begin{abstract}
We show that any optical dissipative structure supported by degenerate optical
parametric oscillators contains a special transverse mode that is free from
quantum fluctuations when measured in a balanced homodyne detection
experiment. The phenomenon is not critical as it is independent of the system
parameters and, in particular, of the existence of bifurcations. This result
is a consequence of the spatial symmetry breaking introduced by the
dissipative structure. Effects that could degrade the squeezing level are considered.

\end{abstract}

\pacs{42.50.Lc, 42.65.Sf}
\maketitle

\textbf{Introduction.} Vacuum quantum fluctuations constitute the ultimate
noise source affecting any coherent radiator, like a laser. These fluctuations
define the so-called standard quantum limit as they set the maximum precision
attainable with classical optical techniques, even rendering the latter
useless in some applications such as precision metrology \cite{pointing} and
quantum information protocols \cite{squeezing1,squeezing2}. It is possible
however to break this limit with the help of quantum states of light: Squeezed
states \cite{squeezing1,squeezing2}, displaying fluctuations below the
standard quantum limit in one of the field quadratures, play a prominent role
in this regard, and are by now routinely generated, e.g., by single-mode
optical parametric oscillators/amplifiers
\cite{pointing,squeezing1,squeezing2,teleportation}.

In the last fifteen years a new branch of quantum optics has emerged under the
general name of "quantum imaging" that studies the spatial aspects of the
field quantum fluctuations and generalizes the study of squeezing to multimode
beams \cite{review}. We consider in this context the squeezing properties of
optical dissipative structures (DS) --transverse patterns--, which are
self-sustained, stable spatial structures that form across the plane
perpendicular to the axis of multi transverse-mode nonlinear resonators
\cite{staliunas}. We consider two outstanding classes of optical DS in the
planar degenerate optical parametric oscillator (DOPO): periodic patterns and
localised structures.

Localised structures of nonlinear optical resonators are also called cavity
solitons (CS) for their resemblance with optical fibre solitons and with
optical spatial solitons. However, in spite of their appealing resemblances,
the latter are Hamiltonian objects resulting from perfect compensation between
dispersion/diffraction and nonlinearity, while CS are also ruled by
dissipation \cite{fauve} and feedback and are not true solitons in the
mathematical sense. The squeezing properties of optical fibre solitons are
well understood since long time ago \cite{solfibra,kozlov}, and those of
optical spatial solitons have been considered more recently \cite{trepslantz}.

We show below that any stationary DOPO DS is phase-squeezed in a special
transverse mode in the linear approximation. Importantly, the degree of
squeezing is independent of the system parameter values and of the existence
of bifurcation points. This result is a direct consequence of the free
diffusion of the DS across the transverse plane, ruled by quantum noise.
Although different in nature, the reported phenomenon resembles the quantum
noise suppresion on the difference of intensities (amplitude-squeezing) of a
two--mode optical parametric oscillator above threshold \cite{reynaudreid},
which is associated to the existence of a continuous diffusion of the phase
difference between the two modes. In our case the fact that squeezing occurs
in a transverse mode with a special spatial shape could be useful for some
applications \cite{pointing}.

\textbf{Model.} We consider the model of \cite{gattiDOPO} for a DOPO with
plane cavity mirrors. A plane wave coherent field of frequency $2\omega
_{\mathrm{s}}$ and amplitude $\mathcal{E}_{\mathrm{in}}$ pumps the resonator
containing a $\chi^{\left(  2\right)  }$ crystal, which converts pump photons
into signal photons (of frequency $\omega_{\mathrm{s}}$) and vice versa. Only
two longitudinal cavity modes, of frequencies $\omega_{0}$ (pump mode) and
$\omega_{1}$ (signal mode), the closest to $2\omega_{\mathrm{s}}$ and
$\omega_{\mathrm{s}}$, respectively, are assumed to be relevant. These modes
are damped at rates $\gamma_{n}$ ($n=0,1$) and losses are assumed to occur at
a single cavity mirror. The intracavity field envelope operators for pump and
signal modes are denoted by $A_{0}\left(  \mathbf{r},t\right)  $ and
$A_{1}\left(  \mathbf{r},t\right)  $, respectively, where $\mathbf{r}=\left(
x,y\right)  $ denotes the transverse coordinates, which obey standard
equal-time commutation relations $\left[  A_{n}\left(  \mathbf{r},t\right)
,A_{n}^{\dag}\left(  \mathbf{r}^{\prime},t\right)  \right]  =\delta\left(
\mathbf{r}-\mathbf{r}^{\prime}\right)  $. Using the positive $P$%
-representation \cite{gattiDOPO,methods,drummond}, which sets a correspondence
between the quantum operators $A_{n}$ and $A_{n}^{\dag}$ and the independent,
stochastic c-number fields $\mathcal{A}_{n}$ and $\mathcal{A}_{n}^{+}$,
respectively, we derive the following model equations \cite{drummond05}:%
\begin{align}
\partial_{t}\mathcal{A}_{0}\left(  \mathbf{r},t\right)   &  =-\gamma
_{0}\left(  1+i\Delta_{0}\right)  \mathcal{A}_{0}+\mathcal{E}_{\mathrm{in}%
}+i\frac{\gamma_{1}}{2}l_{1}^{2}\nabla^{2}\mathcal{A}_{0}\label{dA0}\\
&  -\frac{g}{2}\mathcal{A}_{1}^{2},\nonumber\\
\partial_{t}\mathcal{A}_{1}\left(  \mathbf{r},t\right)   &  =-\gamma
_{1}\left(  1+i\Delta_{1}\right)  \mathcal{A}_{1}+i\gamma_{1}l_{1}^{2}%
\nabla^{2}\mathcal{A}_{1}\label{dA1}\\
&  +g\mathcal{A}_{0}\mathcal{A}_{1}^{+}+\sqrt{g\mathcal{A}_{0}}~\eta\left(
\mathbf{r},t\right)  ,\nonumber
\end{align}
where $\Delta_{0}=\left(  \omega_{0}-2\omega_{\mathrm{s}}\right)  /\gamma_{0}$
and $\Delta_{1}=\left(  \omega_{1}-\omega_{\mathrm{s}}\right)  /\gamma_{1}$
are cavity detunings, $l_{1}=c/\sqrt{2\omega_{\mathrm{s}}\gamma_{1}}$ is a
characteristic (diffraction) length, $c$ is the speed of light in the crystal,
$\nabla^{2}=\partial^{2}/\partial x^{2}+\partial^{2}/\partial y^{2}$ accounts
for diffraction, and $g$ is the (real) coupling coefficient proportional to
the relevant second-order nonlinear susceptibility of the crystal. As
everywhere along the rest of this Letter the equations for $\mathcal{A}_{n}$
are to be complemented by those for the "hermitian conjugate" fields
$\mathcal{A}_{n}^{+}$, which are obtained from those for $\mathcal{A}_{n}$ by
complex-conjugating the parameters and by doing the replacements
$\mathcal{A}_{n}\longleftrightarrow\mathcal{A}_{n}^{+}$ and $\eta
\longleftrightarrow\eta^{+}$. Finally $\eta$ and $\eta^{+}$ are independent,
real white Gaussian noises of zero mean and correlations%
\begin{align}
\left\langle \eta\left(  \mathbf{r},t\right)  \eta\left(  \mathbf{r}^{\prime
},t^{\prime}\right)  \right\rangle  &  =\left\langle \eta^{+}\left(
\mathbf{r},t\right)  \eta^{+}\left(  \mathbf{r}^{\prime},t^{\prime}\right)
\right\rangle \label{noises}\\
&  =\delta\left(  \mathbf{r}-\mathbf{r}^{\prime}\right)  \delta\left(
t-t^{\prime}\right)  .\nonumber
\end{align}
Thus $\mathcal{A}_{n}$ and $\mathcal{A}_{n}^{+}$ are not complex conjugate,
although the stochastic average of any function of them, $\left\langle
f\left(  \mathcal{A}_{0},\mathcal{A}_{0}^{+},\mathcal{A}_{1},\mathcal{A}%
_{1}^{+}\right)  \right\rangle $, yields the normally ordered quantum
expectation value $\left\langle :f\left(  A_{0},A_{0}^{\dag},A_{1},A_{1}%
^{\dag}\right)  :\right\rangle $. For example, the signal photon density is
calculated as%
\begin{equation}
\mathcal{N}_{1}\left(  \mathbf{r},t\right)  =\left\langle A_{1}^{\dag}\left(
\mathbf{r},t\right)  A_{1}\left(  \mathbf{r},t\right)  \right\rangle
=\left\langle \mathcal{A}_{1}^{+}\left(  \mathbf{r},t\right)  \mathcal{A}%
_{1}\left(  \mathbf{r},t\right)  \right\rangle . \label{N1}%
\end{equation}

In the limit of large pump detuning ($\gamma_{0}\left\vert \Delta
_{0}\right\vert \gg\gamma_{1}\left\vert \Delta_{1}\right\vert ,\gamma
_{0},\gamma_{1}$) pump diffraction plays a negligible role and pump fields can
be adiabatically eliminated \cite{DOPOpgl} as $\mathcal{A}_{0}=\gamma
_{1}\left(  \mu+i\frac{\sigma}{\kappa^{2}}\mathcal{A}_{1}^{2}\right)  /g$,
where $\kappa=\gamma_{1}\sqrt{2\left\vert \Delta_{0}\right\vert }/g$,
$\mu=g\left\vert \mathcal{E}_{\mathrm{in}}\right\vert /\left(  \gamma_{1}%
^{2}\left\vert \Delta_{0}\right\vert \right)  >0$ is the dimensionless pump
parameter, $\sigma=\operatorname{sign}\Delta_{0}$, and we wrote $\mathcal{E}%
_{\mathrm{in}}=i\sigma\left\vert \mathcal{E}_{\mathrm{in}}\right\vert $
without loss of generality. Then eq. (\ref{dA1}) becomes
\begin{align}
\partial_{t}\mathcal{A}_{1}  &  =\gamma_{1}\left[  -\left(  1+i\Delta
_{1}\right)  \mathcal{A}_{1}+\mu\mathcal{A}_{1}^{+}+il_{1}^{2}\nabla
^{2}\mathcal{A}_{1}+i\frac{\sigma}{\kappa^{2}}\mathcal{A}_{1}^{2}%
\mathcal{A}_{1}^{+}\right] \label{pgl}\\
&  +\sqrt{\gamma_{1}\left(  \mu+i\frac{\sigma}{\kappa^{2}}\mathcal{A}_{1}%
^{2}\right)  }~\eta.\nonumber
\end{align}

\textbf{Classical dissipative structures.} When noises are ignored and
$\mathcal{A}_{i}^{+}$ is identified with $\mathcal{A}_{i}^{\ast}$ (classical
limit) eq. (\ref{pgl}) coincides with that for a classical DOPO with large
pump detuning \cite{DOPOpgl}, which supports different types of steady and
stable DS, both periodic patterns \cite{longhiapl} or CS
\cite{DOPOpgl,barashenkov2D} depending on the parameter region. All these DS
have the form:%
\begin{gather}
\mathcal{A}_{1}\left(  \mathbf{r}\right)  =\mathcal{\bar{A}}_{1}\left(
\mathbf{r}-\mathbf{r}_{1}\right)  ,\ \ \mathcal{\bar{A}}_{1}\left(
\mathbf{r}\right)  =\kappa e^{i\sigma\theta}\mathcal{F}\left(  \mathbf{r}%
\right)  ,\label{ds}\\
e^{2i\sigma\theta}=\mu^{-1}\left(  1+i\sigma\sqrt{\mu^{2}-1}\right)  ,\ \ \ \\
\left(  \sigma l_{1}^{2}\nabla^{2}-\beta^{2}+\mathcal{F}^{2}\right)
\mathcal{F}=0,\ \beta^{2}=\sigma\Delta_{1}+\sqrt{\mu^{2}-1},\ \ \ \label{beta}%
\end{gather}
where $\mathbf{r}_{1}=\left(  x_{1},y_{1}\right)  $ is arbitrary due to the
translation invariance and $\mathcal{F}$ is real. Note that the only
parameters defining the DS are $\left\{  \sigma,\mu,\Delta_{1}\right\}  $ as
$\kappa$ and $l_{1}$ merely act as scale factors.

\textbf{Dynamics of quantum fluctuations.} Fluctuations around any classical
DS, eq. (\ref{ds}), are studied by setting%
\begin{align}
\mathcal{A}_{1}\left(  \mathbf{r},t\right)   &  =\mathcal{\bar{A}}_{1}\left(
\mathbf{r}-\mathbf{r}_{1}\right)  +a_{1}\left(  \mathbf{r}-\mathbf{r}%
_{1},t\right)  ,\qquad\\
\mathcal{A}_{1}^{+}\left(  \mathbf{r},t\right)   &  =\mathcal{\bar{A}}%
_{1}^{\ast}\left(  \mathbf{r}-\mathbf{r}_{1}\right)  +a_{1}^{+}\left(
\mathbf{r}-\mathbf{r}_{1},t\right)  ,
\end{align}
where, given the translation invariance, the position of the classical DS,
$\mathbf{r}_{1}\left(  t\right)  $, is let to vary in time as it is an
undamped variable that is excited by noise. Expressing the fluctuations as
$\mathbf{a}=\left(  a_{1},a_{1}^{+}\right)  ^{\mathrm{T}}$ ($\mathrm{T}$
denotes transposition) and linearizing the resulting eq. (\ref{pgl}),
stochastic equations for the quantum fluctuations are obtained:%
\begin{equation}
-\kappa\left(  \mathbf{G}_{x}\mathrm{d}x_{1}/\mathrm{d}t+\mathbf{G}%
_{y}\mathrm{d}y_{1}/\mathrm{d}t\right)  +\partial_{t}\mathbf{a}=\gamma
_{1}\mathcal{L}\mathbf{a}+\sqrt{\gamma_{1}}~\mathbf{h}, \label{dirac}%
\end{equation}
where%
\begin{gather}
\mathbf{G}_{j}=\left(  G_{j},G_{j}^{\ast}\right)  ^{\mathrm{T}},\qquad
G_{j}=e^{i\sigma\theta}\partial_{j}\mathcal{F},\qquad j=x,y, \label{Goldstone}%
\\
\mathbf{h}=\left(  \sqrt{\bar{\alpha}_{0}}~\eta,\sqrt{\bar{\alpha}_{0}^{\ast}%
}~\eta^{+}\right)  ^{\mathrm{T}},\qquad\bar{\alpha}_{0}=\mu+i\frac{\sigma
}{\kappa^{2}}\mathcal{\bar{A}}_{1}^{2}\\
\mathcal{L}=\left[
\begin{array}
[c]{cc}%
-1+iL_{1} & \bar{\alpha}_{0}\\
\bar{\alpha}_{0}^{\ast} & -1-iL_{1}%
\end{array}
\right]  ,\ \label{L}\\
\mathcal{L}^{\dag}=\left[
\begin{array}
[c]{cc}%
-1-iL_{1} & \bar{\alpha}_{0}\\
\bar{\alpha}_{0}^{\ast} & -1+iL_{1}%
\end{array}
\right]  ,
\end{gather}
with $L_{1}=\left(  l_{1}^{2}\nabla^{2}-\Delta_{1}\right)  $. The spectra of
the linear operators $\mathcal{L}$ and $\mathcal{L}^{\dag}$, which we
introduce as $\mathcal{L}\mathbf{v}_{i}=\lambda_{i}\mathbf{v}_{i}$,
$\mathcal{L}^{\dag}\mathbf{w}_{i}=\lambda_{i}^{\ast}\mathbf{w}_{i}$, are
clearly relevant \cite{index}. In order to deal with stable DS, it is assumed
that $\operatorname{Re}\lambda_{i}\leq0$ for any $i$ . With the usual
definition of scalar product $\left\langle \mathbf{b}\right\vert
\mathbf{c}\rangle\equiv\int\mathrm{d}^{2}r\ \mathbf{b}^{\dag}\left(
\mathbf{r}\right)  \cdot\mathbf{c}\left(  \mathbf{r}\right)  $, the relation
$\left\langle \mathbf{w}_{i}\right\vert \mathcal{L}\mathbf{c}\rangle
=\lambda_{i}\left\langle \mathbf{w}_{i}\right\vert \mathbf{c}\rangle$ holds.
We assume that all eigenvectors are suitably orthonormalised as $\left\langle
\mathbf{w}_{i}\right\vert \mathbf{v}_{j}\rangle=\delta_{ij}$. In general the
spectra must be computed numerically; nevertheless two general properties of
the discrete spectra can be stated: (i) $\mathbf{G}_{x\left(  y\right)  }$ are
Goldstone modes as $\mathcal{L}\mathbf{G}_{x\left(  y\right)  }=0$ (we write
$\mathbf{v}_{1x\left(  1y\right)  }\equiv\mathbf{G}_{x\left(  y\right)  }$ and
denote by $\mathbf{w}_{1x\left(  1y\right)  }$ the associated adjoint
eigenvectors: $\mathcal{L}^{\dag}\mathbf{w}_{1x\left(  1y\right)  }=0$); and
(ii)
\begin{align}
\mathcal{L}^{\dag}\mathbf{w}_{2x\left(  2y\right)  }  &  =-2\mathbf{w}%
_{2x\left(  2y\right)  },\ \mathbf{w}_{2x\left(  2y\right)  }=\left(
w_{2x\left(  2y\right)  },w_{2x\left(  2y\right)  }^{\ast}\right)
^{\mathrm{T}},\qquad\nonumber\\
w_{2x\left(  2y\right)  }  &  =iG_{x\left(  y\right)  }. \label{w2}%
\end{align}
Properties (i) and (ii) occur independently of the set of parameters $\left\{
\sigma,\mu,\Delta_{1}\right\}  $. Property (i) is a mere consequence of the
translational invariance of the problem. Property (ii) is the key for our
analysis. Note that these eigenvectors exist as the classical DS breaks the
spatial symmetry: for a spatially homogeneous solution ($\mathcal{F}%
=\mathrm{cst}$) $G_{x\left(  y\right)  }=0$.

Before solving eq. (\ref{dirac}) we consider its projections onto the
eigenvectors $\mathbf{w}_{1x\left(  1y\right)  }$ and $\mathbf{w}_{2x\left(
2y\right)  }$ introduced above:%
\begin{equation}
\mathrm{d}x_{1}/\mathrm{d}t=-\sqrt{\gamma_{1}}\kappa^{-1}\xi_{1x}%
,\ \mathrm{d}c_{2x}/\mathrm{d}t=-2\gamma_{1}c_{2x}+\sqrt{\gamma_{1}}\xi_{2x},
\label{proj12}%
\end{equation}
where $\xi_{i}\left(  t\right)  =\left\langle \mathbf{w}_{i}\right\vert
\mathbf{h}\rangle$ are noise sources, $c_{2x}\left(  t\right)  =\left\langle
\mathbf{w}_{2x}\right\vert \mathbf{a}\rangle$ is the projection of the
fluctuations onto the eigenmode $\mathbf{w}_{2}$, and corresponding
expressions for $\mathrm{d}y_{1}/\mathrm{d}t$ and $\mathrm{d}c_{_{2y}%
}/\mathrm{d}t$. Note that the equation for $x_{1}$ is diffusive, as
anticipated, because Goldstone modes are excited without cost \cite{zambrini}.
We further notice that the equation for $c_{2x}$ is analogous to that derived
in \cite{gattimancini} for the hexagonal mode stationary phase in a Kerr
cavity, which is later interpreted as the hexagonal pattern transverse
momentum in \cite{gomila}.

\textbf{Squeezing via optical homodyning.} We consider the squeezing
properties of a DS as measured in a balanced homodyne detection experiment
\cite{gattiLOF}: The outgoing quantum field, $A_{1,\mathrm{out}}\left(
\mathbf{r},t\right)  $, is combined in a beam splitter with a local oscillator
field (LOF) that lies in an intense (multimode) coherent state of transverse
complex envelope $\alpha_{\mathrm{L}}\left(  \mathbf{r}-\mathbf{r}%
_{\mathrm{L}}\left(  t\right)  \right)  $, which is allowed to be dynamically
shifted. In the detection of squeezing one measures the normally ordered part
of the fluctuation spectrum of the intensity difference between the two output
ports of the beam splitter, $S$, which can be computed as \cite{gattiLOF}%
\begin{align}
S\left(  \omega\right)   &  =2\gamma_{1}\int_{-\infty}^{+\infty}\mathrm{d}\tau
e^{-i\omega\tau}\langle\delta\mathcal{E}_{\mathrm{H}}\left(  t+\tau\right)
\delta\mathcal{E}_{\mathrm{H}}\left(  t\right)  \rangle,\label{Sw}\\
\delta\mathcal{E}_{\mathrm{H}}\left(  t\right)   &  =\frac{1}{\sqrt
{\int\mathrm{d}^{2}r\left\vert \alpha_{\mathrm{L}}\right\vert ^{2}}}%
\langle\boldsymbol{\alpha}_{\mathrm{L}}\left(  \mathbf{r}+\boldsymbol{\rho
}\left(  t\right)  \right)  \mid\mathbf{a}\left(  \mathbf{r},t\right)
\rangle,\label{dE}\\
\boldsymbol{\alpha}_{\mathrm{L}}  &  =\left(  \alpha_{\mathrm{L}}%
,\alpha_{\mathrm{L}}^{\ast}\right)  ^{T},\ \ \ \ \boldsymbol{\rho}%
=\mathbf{r}_{1}-\mathbf{r}_{\mathrm{L}}.
\end{align}
When $A_{1,\mathrm{out}}\left(  \mathbf{r},t\right)  $ is in a (multimode)
coherent state $S\left(  \omega\right)  =0$, the standard quantum limit. On
the other hand $S\left(  \omega_{\mathrm{s}}\right)  =-1$ signals complete
absence of quantum fluctuations at $\omega=\omega_{\mathrm{s}}$.

Let us assume momentarily that we can set $\boldsymbol{\rho}=0$ in eq.
(\ref{dE}), which means that we can shift the LOF according to the DS
movement. Let us choose a LOF with $\alpha_{\mathrm{L}}=iG_{x\left(  y\right)
}$ (i.e., $\boldsymbol{\alpha}_{\mathrm{L}}=\bm{w}_{2x}$, eq. (\ref{w2})) so
that $\delta\mathcal{E}_{\mathrm{H}}\left(  t\right)  =c_{2x}\left(  t\right)
$, see after eq. (\ref{proj12}). Standard techniques \cite{methods} applied to
eq. (\ref{proj12}) allow to compute the stochastic correlation $\langle
\delta\mathcal{E}_{\mathrm{H}}\left(  t+\tau\right)  \delta\mathcal{E}%
_{\mathrm{H}}\left(  t\right)  \rangle=-\tfrac{1}{2}e^{-2\gamma_{1}\left\vert
\tau\right\vert }$. Finally using eq. (\ref{Sw}) we get%
\begin{equation}
S\left(  \omega\right)  =-\frac{4\gamma_{1}^{2}}{4\gamma_{1}^{2}+\omega^{2}},
\label{S2}%
\end{equation}
which is the main result of this Letter: As $S\left(  \omega=0\right)  =-1$,
DOPO DS display \textit{perfect squeezing} at $\omega=0$ when probed with the
appropriate LOF ($\alpha_{\mathrm{L}}=iG_{x\left(  y\right)  }$). As eq.
(\ref{S2}) is independent of the kind of DS and of the system parameters, the
result is universal and independent of the existence of bifurcations. This LOF
is, in principle, easily realisable as it is the $\pi/2$ phase-shifted
gradient of the corresponding DS envelope, eq. (\ref{ds}), which can be easily
synthesised by, e.g., Fourier filtering.

It is interesting to notice that in \cite{gattimancini}, a perfectly squeezed
spatial mode was identified in the hexagonal pattern arising in a Kerr cavity
and, as in our case, the result is independent of the parameter values.
Although derived by different means from the ones used here, this result is
very likely connected to the one we have just derived.

We note that the linearised approach is valid, in principle, when all
eigenvalues are strictly negative, as then all fluctuations remain small. In
our case however a null eigenvalue exists always --that associated with the
Goldstone mode. Nevertheless that eigenvalue is just the responsible for the
continuous diffusion of the position of the DS (similarly to the continuous
diffusion of the phase difference in \cite{reynaudreid}), and does not entail
an energetic divergence. Hence one can be confident that the linearised theory
developed here represents quite an accurate description and that a nonlinear
treatment \cite{drummond05} would not lead to dramatically different results.

Next we consider two effects that could degrade the measured squeezing level.
Although everything to be said applies to any DOPO DS, we focus on the bright
CS, which exists for $\sigma=+1$ \cite{DOPOpgl,barashenkov2D}, for the sake of clarity.

\textbf{Influence of the CS movement.} Equation (\ref{S2}) is valid if we use
a movable LOF which exactly follows the CS movement. This could be done by
tracking the movement of $A_{0,\mathrm{out}}$, which is correlated with
$A_{1,\mathrm{out}}$, without disturbing the subharmonic CS. The output of
this continuous measurement would be then fed into a positioning system
controlling $\mathbf{r}_{\mathrm{L}}\left(  t\right)  $, giving rise, in
general, to a time delay $t_{\mathrm{d}}$ so that $\mathbf{r}_{\mathrm{L}%
}\left(  t\right)  =\mathbf{r}_{1}\left(  t-t_{\mathrm{d}}\right)  $ yielding
$\boldsymbol{\rho}\left(  t\right)  =\mathbf{r}_{1}\left(  t\right)
-\mathbf{r}_{1}\left(  t-t_{\mathrm{d}}\right)  $. The point is how much the
CS position diffuses in time as compared with its width $\Delta x$. Standard
techniques \cite{methods} applied to eq. (\ref{proj12}) allow to obtain
$\left\langle \boldsymbol{\rho}^{2}\left(  t\right)  \right\rangle
=Dt_{\mathrm{d}}$, where the diffusion constant $D\sim\gamma_{1}\kappa^{-2}$
\cite{diffusion}. Thus $\left\langle \boldsymbol{\rho}^{2}\left(  t\right)
\right\rangle /\Delta x^{2}\sim\gamma_{1}t_{\mathrm{d}}/N_{1}$, being
$N_{1}\sim\left(  \kappa l_{1}\right)  ^{2}$ the number of intracavity signal
photons in one CS and $\Delta x\sim l_{1}$ its width \cite{CSproperties}. We
see that $N_{1}$ acts as an inertial mass. Using realistic values for the
system parameters \cite{parameters} one has $N_{1}\sim10^{12}$, and
$\sqrt{\left\langle \boldsymbol{\rho}^{2}\left(  t\right)  \right\rangle
}/\Delta x\lesssim5\cdot10^{-4}$ for a delay time $t_{\mathrm{d}}%
\lesssim1\mathrm{ms}$. Thus the relative error existing between the location
of the CS center and that of the LOF is very small as compared with the CS
width. In order to assess the negligible influence of this effect we expand
eq. (\ref{dE}) up to second order in $\boldsymbol{\rho}$. The squeezing
spectrum is then given by eq. (\ref{S2}) plus a correction proportional to
$\left\langle \boldsymbol{\rho}^{2}\left(  t\right)  \right\rangle /\left(
\Delta x\right)  ^{2}\lesssim3\cdot10^{-7}$, again for $t_{\mathrm{d}}%
\lesssim1\mathrm{ms}$, which is absolutely negligible.

Let us now consider how the squeezing properties of the CS are modified if the
LOF is kept fixed, which corresponds to a simpler scheme. Due to the unbounded
movement of the soliton, one must perform the heterodyning experiment in a
short time (call it $t_{\mathrm{H}}$) in order to obtain significant
squeezing. If we assume that at $t=0$ the LOF and the CS centers are made to
coincide one has $\boldsymbol{\rho}\left(  t\right)  =\mathbf{r}_{1}\left(
t\right)  -\mathbf{r}_{1}\left(  0\right)  $ and $\sqrt{\left\langle
\boldsymbol{\rho}^{2}\left(  t\right)  \right\rangle }/\Delta x\sim
\sqrt{\gamma_{1}t/N_{1}}\sim10^{-6}\sqrt{\gamma_{1}t}$. Then, if we take
$t_{\mathrm{H}}\lesssim1\mathrm{ms}$ (for the used parameters) we can again
take $\boldsymbol{\rho}=0$ in eq. (\ref{dE}), which yields $\langle
\delta\mathcal{E}_{\mathrm{H}}\left(  t+\tau\right)  \delta\mathcal{E}%
_{\mathrm{H}}\left(  t\right)  \rangle=-\frac{1}{2}e^{-2\gamma_{1}\left\vert
\tau\right\vert }$ as before. The squeezing spectrum is then given by eq.
(\ref{Sw}) with the limits of integration being replaced by $\mp
\frac{t_{\mathrm{H}}}{2}$. The result reads as eq. (\ref{S2}) plus a
correction proportional to $e^{-\gamma_{1}t_{\mathrm{H}}}$, which is virtually
zero. Then the obtention of almost optimal levels of squeezing is not affected
in practice by the existing CS movement. We note that this insensitivity
contrasts with the issue of quantum images \cite{lugiatoquantim}, whose
squeezing properties are washed out by their jittering. This is a consequence
of the strong inertia ($\propto N_{1}^{1/2}$) that the CS movement displays
against fluctuations, as compared with the below- or close-to-threshold
emission analysed in \cite{lugiatoquantim}.%
\begin{figure}
[ptb]
\begin{center}
\includegraphics[
trim=0.996409in 3.638446in 1.359822in 5.081718in,
height=3.2053cm,
width=8.4751cm
]%
{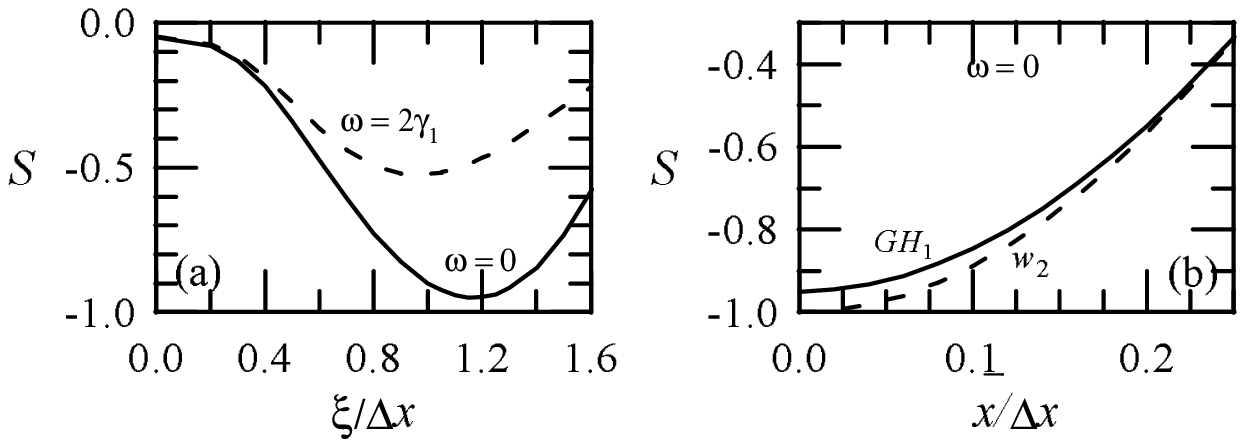}%
\caption{Squeezing level (at the labeled frequencies) displayed by the 1D CS
when nonideal LOFs are used. In (a) a Gauss-Hermite LOF ($GH_{1}$) of width
$\xi$ is used. In (b) LOFs displaced $\bar{x}$ from the CS center are
considered ($w_{2}$ denotes a special LOF, see text). $\Delta x=l_{1}/\beta$
denotes the CS width. Parameters are $\sigma=+1$, $\Delta_{1}=1$, $\mu=1.2$.}%
\end{center}
\end{figure}

\textbf{Influence of the shape and positioning of the LOF.} Up to here we have
dealt with a special LOF. However the use of a LOF with the exact form or
precisest positioning is not critical as we show next. In what follows we
ignore the negligible influence of the CS movement and consider a static LOF.
The study requires fully solving eq. (\ref{dirac}), what was done by using the
(biorthogonal) basis $\left\{  \mathbf{v}_{i},\mathbf{w}_{j}\right\}  $ formed
by the eigenvectors of the linear operators $\mathcal{L}$ and $\mathcal{L}%
^{\dag}$ in eq. (\ref{L}) \cite{base}. This technique is highly convenient as
it allows to circumvent the numerical simulation of the Langevin eq.
(\ref{dirac}), which can be problematic and, in any case, extremely time
consuming. We write the field fluctuations and the LOF vector as
$\mathbf{a}\left(  \mathbf{r},t\right)  =\sum c_{i}\left(  t\right)
\mathbf{v}_{i}\left(  \mathbf{r}\right)  $, $c_{i}=\left\langle \mathbf{w}%
_{i}\right\vert \mathbf{a}\rangle$, where the expansion excludes the Goldstone
modes as $\mathbf{r}_{1}$ is to denote the position of the CS. By substituting
this expansion into eq. (\ref{dirac}) and projecting, one obtains
$\mathrm{d}c_{i}/\mathrm{d}t=\gamma_{1}\lambda_{i}c_{i}+\sqrt{\gamma_{1}%
}\left\langle \mathbf{w}_{i}\right\vert \mathbf{h}\rangle$. The study is
further facilitated by expressing a general LOF as $\boldsymbol{\alpha
}_{\mathrm{L}}=\sum\alpha_{i}\mathbf{w}_{i}$, $\alpha_{i}=\left\langle
\mathbf{v}_{i}\right\vert \boldsymbol{\alpha}_{\mathrm{L}}\rangle$. One
obtains%
\begin{align}
S\left(  \omega\right)   &  =\frac{2\gamma_{1}}{\sqrt{\int\mathrm{d}%
^{2}r\left\vert \alpha_{\mathrm{L}}\right\vert ^{2}}}\sum_{p,q}\alpha
_{p}^{\ast}\alpha_{q}S_{p,q}\left(  \omega\right)  ,\ \ \\
S_{p,q}\left(  \omega\right)   &  =\int_{-\infty}^{+\infty}\mathrm{d}\tau
e^{-i\omega\tau}\left\langle c_{p}\left(  t+\tau\right)  c_{q}\left(
t\right)  \right\rangle \ ,\nonumber
\end{align}
which are easily evaluated by using standard methods \cite{methods}. In the
end all we need is to compute the spectra of $\mathcal{L}$\ and $\mathcal{L}%
^{\dag}$, which was done by using a Fourier method \cite{alexeeva}. We limit
here our study to the 1D CS ($\mathcal{F}=\sqrt{2}\beta\operatorname{sech}%
\left(  \beta x/l_{1}\right)  $ \cite{DOPOpgl}) for the sake of computational
economy. The influence of the LOF shape was studied for a Gauss-Hermite mode
of appropriate phase, $\alpha_{\mathrm{L}}\left(  x\right)  =GH_{1}\left(
x\right)  =ie^{i\theta}x~e^{-\frac{1}{2}\left(  x/\xi\right)  ^{2}}$, which is
similar to $w_{2}=iG_{x}=ie^{i\theta}\partial_{x}\mathcal{F}$. Figure 1(a)
shows that quite high levels of squeezing can be reached even with this non
ideal LOF. Finally the influence of a mispositioning was studied both for
$\alpha_{\mathrm{L}}\left(  x\right)  =w_{2}\left(  x-\bar{x}\right)  $ and
$\alpha_{\mathrm{L}}\left(  x\right)  =GH_{1}\left(  x-\bar{x}\right)  $, fig.
1(b): Mispositionings as high as 15\% of the CS width still yield quite good
levels of squeezing as well.

In summary we have shown that optical dissipative structures sustained by a
DOPO always contain a transverse mode that is completely free from
zero-frequency quantum fluctuations. Unlike single-mode cavity squeezing,
which is perfect only at bifurcation points, our result does not depend on the
parameter setting.

\begin{acknowledgments}
We gratefully acknowledge fruitful discussions with A. Gatti, K. Staliunas and
J.A. de Azc\'{a}rraga. We are grateful to V.V. Kozlov for making ref.
\cite{kozlov} available to us. This work has been supported by the spanish
Ministerio de Educaci\'{o}n y Ciencia and the European Union FEDER through
Projects BFM2002-04369-C04-01 and FIS2005-07931-C03-01.
\end{acknowledgments}


\begin{thebibliography}{99}
\bibitem {index}The index $i$ accounts for both the discrete and the
continuous spectra. In the latter case, Kroenecker deltas and sums should be
understood as containing suitable Dirac deltas and integrals.
\bibitem {diffusion}$D=2\gamma_{1}\kappa^{-2}\operatorname{Re}\int
\mathrm{d}^{2}r\ \left[  w_{1x}^{2}\left(  \mathbf{r}\right)  +w_{1y}%
^{2}\left(  \mathbf{r}\right)  \right]  $\ $\left[  \mu+i\sigma\kappa
^{-2}\mathcal{\bar{A}}_{1}^{2}\left(  \mathbf{r}\right)  \right]  $
$e^{-2i\sigma\theta}$. The integral is dimensionless and is found numerically
to be of order unity.
\bibitem {CSproperties}The intracavity signal photon density is given by eq.
(\ref{N1}). When fluctuations are ignored $\mathcal{N}_{1}=\left\vert
\mathcal{\bar{A}}_{1}\right\vert ^{2}=\kappa^{2}\mathcal{F}^{2}$, eq.
(\ref{ds}). The number of intracavity signal photons in a CS is thus
$N_{1}=2\pi\int\mathrm{d}r\,r\,\mathcal{N}_{1}$. The area of the CS is
$\Sigma\equiv2\pi\int\mathrm{d}r\,r\,\left(  \pi r^{2}\right)  \mathcal{N}%
_{1}/N_{1}$. By integrating the equation that determines $\mathcal{F}$, eq.
(\ref{beta}) \cite{barashenkov2D}, we obtain $N_{1}\simeq11.70\left(  \kappa
l_{1}\right)  ^{2}$ and $\Sigma\simeq3.731\left(  l_{1}/\beta\right)  ^{2}$.
The CS width is thus $\Delta x\sim\sqrt{\Sigma}\sim l_{1}/\beta\sim l_{1}$ as
$\beta$, eq. (\ref{beta}), is of order unity \cite{barashenkov2D}.
\bibitem {parameters}In 2D geometries $g=\frac{3}{2}\chi^{\left(  2\right)
}\left(  2\pi n\right)  ^{-3}\sqrt{\hslash/\varepsilon_{0}L_{\mathrm{z}}%
}\omega_{\mathrm{s}}^{3/2}$, being $\chi^{\left(  2\right)  }$ the relevant
nonlinear susceptibility, $n$ the refractive index of the crystal, and
$L_{\mathrm{z}}$ the thickness of the crystal along the resonator axis. We
take $\chi^{\left(  2\right)  }=2\cdot10^{-11}\mathrm{m}\cdot\mathrm{V}^{-1}$,
$\omega_{\mathrm{s}}=1.885\cdot10^{15}\mathrm{s}^{-1}$ (signal wavelength of
$1\mathrm{\mu m}$), $n=2.5$, $L_{z}=10^{-2}\mathrm{m}$, $\gamma_{0}=\gamma
_{1}=3\cdot10^{8}\mathrm{s}^{-1}$, and $\delta_{0}=10\gamma_{0}$.
\bibitem {base}The fact that $\left\{  \mathbf{v}_{i},\mathbf{w}_{j}\right\}
$ form a basis was assessed by numerical diagonalization of $\mathcal{L}$ and
$\mathcal{L}^{\dag}$ \cite{alexeeva}: The number of independent eigenvectors
($n_{\mathrm{e}}$) so obtained was found to be equal to the number
($n_{\mathrm{m}}$) of Fourier modes used to discretize the problem. As check
of the method we applied the same technique to the related problem of
(conservative) nonlinear Schr\"{o}dinger (NLS) solitons, whose associated
linear operators are similar to $\mathcal{L}$ and $\mathcal{L}^{\dag}$
\cite{kozlov}, obtaining $n_{\mathrm{e}}<n_{\mathrm{m}}$, in agreement with
the well known result that now the eigenvectors do not form a basis
\cite{kozlov}.
\end{thebibliography}
\end{document}